\documentclass[preprint2]{aastex}

\newcommand{\etal}{{et~al.}}

\shorttitle{Polarization of GRB\,990712}
\shortauthors{Rol~\etal}

\begin{document}

\title{GRB\,990712: First Indication of Polarization Variability in a
       Gamma-ray Burst Afterglow\thanks{Based on observations with the
       Very Large Telescope at the European Southern Observatory,
       Chile. Program 63.O-0567(A)}}

\author{E. Rol\altaffilmark{1}, R. A. M. J. Wijers\altaffilmark{2},
P. M. Vreeswijk\altaffilmark{1}, L. Kaper\altaffilmark{1},
T. J. Galama\altaffilmark{3}, {J. van Paradijs\altaffilmark{1,
4}}$\ ^{\thanks{deceased}}$, C. Kouveliotou\altaffilmark{5, 6},
N. Masetti\altaffilmark{7}, E. Pian\altaffilmark{7},
E. Palazzi\altaffilmark{7}, F. Frontera\altaffilmark{7,8},
E.P.J. van den Heuvel\altaffilmark{1}}

\altaffiltext{1}{Astronomical Institute 'Anton Pannekoek', University
of Amsterdam, Kruislaan 403, 1098 SJ Amsterdam; evert@astro.uva.nl}
\altaffiltext{2}{SUNY Stony Brook, Department Physics and Astronomy,
Stony Brook NY 11794-3800 USA} 
\altaffiltext{3}{California Institute of Technology, 1200 East
California Boulevard, Pasadena, California 91125}
\altaffiltext{4}{Physics Department, University of Alabama in
Huntsville, Huntsville, AL 35899} 
\altaffiltext{5}{NASA Marshall Space Flight Center, Code
ES-84, Huntsville, AL 35812} 
\altaffiltext{6}{Universities Space Research Association}
\altaffiltext{7}{ITESRE-CNR Bologna,
Via P. Gobetti 101, 40129 Bologna, Italy}
\altaffiltext{8}{Physics Department, University of Ferrara,
Via Paradiso 12, 44100 Ferrara, Italy}

\begin{abstract}

We report the detection of significant polarization in the optical
afterglow of GRB\,990712 on three instances 0.44, 0.70 and
1.45 days after the gamma-ray burst, with ($P$, $\theta$) being
($2.9\% \pm 0.4\%$, $121.1\degr \pm 3.5\degr$), ($1.2\% \pm 0.4\%$, $116.2\degr
\pm 10.1\degr$) and ($2.2\% \pm 0.7\%$, $139.2\degr \pm 10.4\degr$)
respectively. The polarization is intrinsic to the afterglow. The
degree of polarization is not constant, and smallest at the second
measurement.  The polarization angle does not vary significantly
during these observations. We find that none of the existing models
predict such polarization variations at constant polarization angle,
and suggest ways in which these models might be modified to
accommodate the observed behavior of this afterglow.

\end{abstract}


\section{Introduction \label{intro}}

The radiation from gamma-ray burst afterglows has been hypothesized
from early on to be synchrotron emission from relativistic electrons
\citep[e.g.,][]{Rees92,Meszaros97}. The observation of broad-band
afterglow spectra \citep{Galama98B,Bloom98} and the fact that the temporal
evolution of the flux from radio to X-ray wavelengths often follows
simple model predictions support this theoretical framework
\citep{Sari00,piran99,Paradijs00}. Since synchrotron radiation is intrinsically
highly polarized (with degrees of linear polarization up to
60--70\%, \citealt{Rybicki79}), it would be natural to expect
polarization in afterglow emission.

The first attempt to measure optical polarization of afterglow
emission, for GRB\,990123, set an upper limit of 2.3\%
\citep{Hjorth99A}.  The next attempts to measure polarization were
successful, leading to the detection of polarization of 1.7\% in the
afterglow of GRB\,990510 \citep{Covino99A,Wijers99A}. Such low
percentages of polarization are somewhat surprising, and models have
been mostly concerned with trying to explain the {\it depolarization}
of gamma-ray bursts. Two principal hypotheses have been
advanced. First, that the emission from the GRB is so highly ordered
that the symmetry of the source causes the net polarization to average
out to zero, even though it is locally high everywhere
\citep{Medvedev99}.  Second, that the emitting medium is very random
on small scales, and that the emission we see is composed of many
uncorrelated polarization patches so that the mean is again close to
zero \citep{Gruzinov99B}. Both these models easily obtain the low
polarization levels observed, and they differ only on the
circumstances under which polarization may be observed, and what its
temporal evolution should be. In the afterglow of GRB\,990510, the
close spacing of the early observations and the large measurements
errors later on precluded any strong conclusions about preferred
models.

GRB\,990712 was detected with the \emph{Beppo}SAX Gamma-Ray Burst
Monitor and Wide Field Cameras Unit 2 on 1999 July 12.69655 UT
\citep{Heise99C}. The first optical follow-up observations started 4.16
hours after the burst, leading to the discovery of an optical
transient (OT) of magnitude $R_{OT} \approx 19.4$ \citep{Bakos99A}. The
light curve of the OT is quickly dominated by the light of its fairly
bright host galaxy, with $R_H \approx 22$.

A detailed description of the photometry of the OT is given by
\citet{sahu00} and \citet{Hjorth00A}.  \citet{Galama99C} determined a
redshift of $z=0.430\pm0.005$ for the OT. Spectroscopic observations
of GRB\,990712 are reported in \citet{Hjorth00A} and \citet{Vreeswijk00}.

The outline of this paper is as follows: in
Sect.~\ref{observations} we describe the observations and data
reduction. We analyze the results in Sect.~\ref{results} and discuss
their significance for existing models in Sect.~\ref{discussion}, as
well as possible modifications to those models. Finally, we
summarize our findings in Sect.~\ref{conclusions}.

\section{Observations \label{observations}}

Three epochs of observations were taken on July 13 and 14 with the
Very Large Telescope (VLT) of the European Southern Observatory at
Paranal, Chile, using VLT Unit Telescope 1 (\emph {Antu}) and the FOcal Reducer
low dispersion Spectrograph (FORS1). All images were taken with a
Bessel $R$ band filter.  In order to obtain the degree of linear
polarization, a Wollaston prism and a half wavelength phase retarder
plate were used as polarization optics. The Wollaston prism separates
the incident light into two components (an ordinary and an
extra-ordinary component), where the half wavelength plate is used to
determine which Stokes parameter is measured ($U$ or
$Q$). A mask producing 22\arcsec\ wide parallel strips was used
to avoid overlap of the ordinary and extra-ordinary rays.  Each
observation consisted of four exposures centered on the
position of the OT, with the phase retarder plate rotated by
$22.5\degr$ between successive exposures.

\setcounter{footnote}{0}
The data were reduced in a standard way with the \emph {NOAO
IRAF\footnote{IRAF is distributed by the National Optical Astronomy
Observatories, which are operated by the Association of Universities
for Research in Astronomy, Inc., under cooperative agreement with the
National Science Foundation.}}  package {\emph {ccdred}}, first
bias-subtracted and then flat-fielded. The fluxes of the point sources
in the field were determined using aperture photometry, with an
aperture radius of one FWHM of the Point Spread Function.

To correct for any instrumental or local interstellar polarization, we
measure the polarization of the OT relative to 21 field stars in the
same image. We have plotted the Stokes parameters $U$ and $Q$ of
both the field stars and the OT in Figure \ref{figure:quplot}. The OT
clearly stands out, having the lowest $U$ value.  We verified that
there are no systematic variations of $U$ and $Q$ of the
field stars with magnitude or position on the CCD, and therefore we
can correct the polarization parameters of the OT for foreground
effects by subtracting the mean $U$ and $Q$ values of the
field stars from those of the OT.

\section{Results \label{results}}

To derive the degree of linear polarization, we first calculate $Q$
and $U$ using standard equations \citep[see, e.g.,][]{Ramaprakash98},
both for the OT and for the field stars. We then subtract the average
$Q$ and $U$ values of the field stars from those of the OT. The
resulting $Q$ and $U$ are used to calculate the degree and the
position angle of the linear polarization. We have also used a
different method \citep[see, e.g.,][]{Serego97}, which within the
errors leads to the same results. This method uses the relation
$S(\phi) = P \cos 2 ( \theta - \phi)$, with the parameter $S(\phi)$
being a measure for the ratio between the two components of the
incident light, separated by the Wollaston prism, and $\phi$ the
corresponding angle of the prism. Figure \ref{figure:cosfit} gives
a cosine fit to the data of the first epoch.

In Figure \ref{figure:lc-pol} we have plotted the polarized
flux together with the R- and V-band light curve. The plot clearly shows the
change in polarization, while the light curve exhibits a smooth
decline. There is also no indication of a change in the (V-R) color of
the afterglow.

We assume that the degree of linear polarization and the position
angle are constant during the observation. The total intensity of the
transient is not, as it is decaying according to a power-law, $I
\propto t^{-\alpha}$, with $\alpha = 0.97$ \citep{sahu00}. The fact
that the powerlaw index is known gives us the possibility to correct
for the decaying intensity. These corrections turn out to be so small
that we don't find any differences in the polarization between an
assumed constant intensity from the start to the end of the
observation, and a powerlaw-like declining intensity.

Both the degree of linear polarization and the position angle versus
the time since the burst are plotted in Figure
{\ref{figure:ptheta-plot}.  We see that the polarization percentage,
$P$, decreases between 0.44 and 0.7 days after the burst, with
3.2$\sigma$ significance. Then at 1.45 days, $P$ is greater again, but
since the difference between the last and middle observation is only
1.5$\sigma$, we cannot be too sure about this rise of $P$. 

In case that the degree of polarization would be
constant, with a mean value of $2.1\% \pm 0.3\%$, the $P$
value at epoch 1 is just within 3$\sigma$ above the mean, where for
epoch 2 it is just 3$\sigma$ below it. The variability could then be
caused by a systematic error. To check this, we have compared the
polarization values of the separate field stars at each of the three
epochs with the mean value. We have not found any
evidence for such a systematic error, and conclude that the observed
variability in the polarization is most likely intrinsic to the source.

Therefore,
we think that we have clearly detected, for the first time, variation
in the polarization of a GRB afterglow, which could either be a
decline that is initially steep and then levels off, or a decline
followed by a rise. The polarization angle, at the same time, never
changed by more than one standard deviation from one observation to
the next.

As argued for the case of GRB\,990510, polarization in a rapidly
varying source is unlikely to be induced by interstellar scattering
\citep{Wijers99A}. In the present case, where the polarization itself
varies, this is even more strongly so: stars with polarization induced
by interstellar scattering are favorite polarization standards,
because their degree of polarization is very constant.

\section{Discussion \label{discussion}}

Intrinsic polarization from a synchrotron source can be as large as
$P_{max} \sim$ 60--70\% \citep{Rybicki79} if the magnetic field is
oriented in one direction. However, for an unresolved source the net
polarization will be small if the different directions of the
polarization average out. This could be caused by highly tangled
magnetic fields, or by a very simple symmetry in the large-scale field
pattern.

If the magnetic field in the GRB afterglow is highly tangled, we can
think of it as a source consisting of N patches \citep{Gruzinov99B}.
The net polarization resulting from the source will then be of order
${P_{max}}/\sqrt{N}$.  The maximum degree of linear polarization
observed (at epoch one) requires a magnetic field divided up into
$\sim 400$ patches. For the second epoch, we see a decrease in the
degree of linear polarization (with $\theta$ being constant), which
would lead to $\sim 2500$ patches, six hours later.  We note, however,
that in that case the polarization angle should also vary by the same
percentage, implying that we expect a change of order a radian between
the first and second epoch in a tangled-field model. However, we see
no significant change in the polarization angle, with a 1-radian
variation ruled out at the 4$\sigma$ level, making this model
unlikely.

A symmetric model for the polarization cancellation arises when
the magnetic field is ordered perpendicular to or parallel
to the ring of emission we see from the afterglow at any
given time \citep{Waxman97B,Panaitescu98A,Sari98A}. This could
naturally arise in some instabilities that generate magnetic fields
\citep{Medvedev99,Gruzinov99C}. In a spherically symmetric
fireball, the polarization would then be exactly zero, so an extra
effect is needed to break the symmetry and get a net polarization. One
possibility is that some turbulence induces brightness variations,
thus weighting some polarization directions more, or that an external
effect such as scintillation or microlensing might enhance the emission
from some parts of the ring \citep{Loeb98,Medvedev99}.  However, any
polarization variations from such a mechanism would be expected to be
both in degree and angle, contrary to what we see.

Another symmetric model is a jet: as explained in \cite*{Ghisellini99}
and \cite*{Sari99}, a collimated burst will naturally exhibit
polarization, up to 20\%, around the time when the light curve
steepens. However, we have no evidence of such a break in the light
curve of GRB\,990712. Also, in a jet model, the degree of polarization
varies without change of the polarization angle, until the
polarization goes through zero and the angle suddenly changes by 90
degrees. Therefore, one could in principle have a situation of varying
polarization without variation of the angle as we see here in a jet
model. However, the sharp drop from epoch 1 to 2 followed by little
change to epoch 3 seems hard to get without a zero transit around
epoch 2 if we compare Figure \ref{figure:ptheta-plot} with the
theoretical curves of \citet[][Fig. 4]{Sari99} and
\citet[][Fig. 4]{Ghisellini99}. So at least for the simple models
published thus far, our data cannot be explained by a beamed jet.

In summary, current models of polarization variations predict
significant changes of the polarization angle along with variations of
the degree of polarization, or the degree of polarization to be around
a peak for a constant polarization angle in case of a jet model. Since
we do not observe such a combination of the polarization angle and
degree of polarization in the afterglow of GRB\,990712, we conclude
that no current model adequately explains our data.

\section{Conclusions \label{conclusions}}

We have observed significant polarization for the afterglow of
GRB\,990712 during three epochs from 0.4 to 1.5 days after the burst.
The polarization percentage varies by 3.2$\sigma$ from 3.0\% at the first
epoch to 1.2\% at the second, while the polarization angle remains
essentially constant. We show that neither tangled-field nor
broken-symmetry models of the polarization of afterglows can explain
this constancy of the polarization angle while the degree of
polarization goes up and down. It appears that the afterglow
polarization has some amount of memory for direction, while varying in
strength. One might speculate that this is telling us something about
the formation of the magnetic field in the shocked afterglow
material. Possibly, this field is grown from a seed field that is
embedded in the swept-up ambient material. While the amount of
amplification and field strength may vary with time, and give rise to
variable polarization, the direction of the seed may remain imprinted
on the amplified field, and thus the polarization angle may be
relatively constant. (Few-day old afterglows are $10^{16}-10^{17}$\,cm
in size, on which scale interstellar magnetic fields can be coherent.)

The results show that in order to further test models, we need more
measurements per burst, to better sample the polarization behavior,
and possibly over a larger range of time. Since the measurements
require photometry with a S/N of 300 or so, they already stretch the
capabilities of the VLT after 1.5--2 days, so it is unlikely that we
shall be able to do measurements of the polarization at later
times. However, it may be possible to extend the time interval of
polarization measurements to earlier times. In a power-law process
like GRB afterglows, it may be as profitable to move the first
observing time forward from 0.4 to 0.2 days as it is to extend the
last one from 2 to 4 days.

\acknowledgments We thank the ESO Paranal Observatory staff for
carrying out the observations. ER is supported by NWO grant nr.
614-51-003. PMV is supported by the NWO Spinoza grant. LK is supported
by a fellowship of the Royal Academy of Sciences in the
Netherlands. TJG acknowledges support from the Sherman Fairchild
Foundation. We thank the referee for useful suggestions.


\begin{figure}
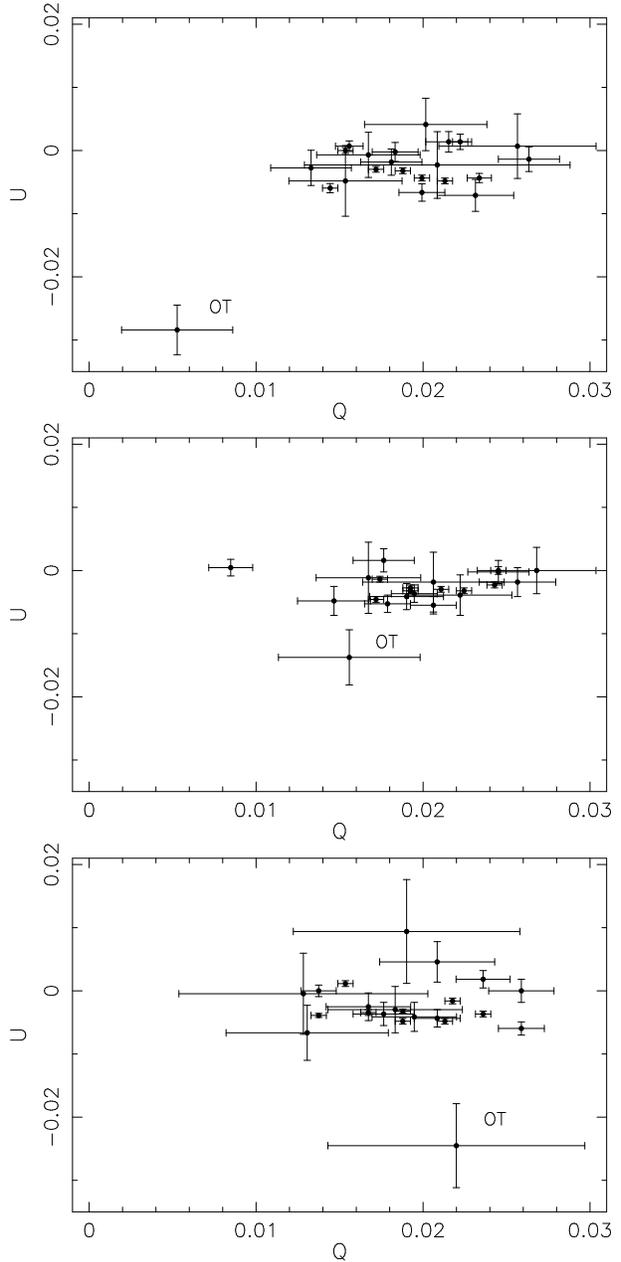

\includegraphics[angle=-90,width=8cm]{fig1a.eps}\\
\includegraphics[angle=-90,width=8cm]{fig1b.eps}\\
\includegraphics[angle=-90,width=8cm]{fig1c.eps}
\caption{\label{figure:quplot}($Q$, $U$) plot of the stars in the
field after instrumental correction, for epoch 1, 2, and 3 (top,
middle and bottom respectively). The OT clearly stands out with respect
to the other stars in the field, indicating that it has a
considerable intrinsic amount of polarization. ($Q$, $U$) averages
of the field stars for the three epochs are ($0.0194 \pm 0.0004$,
$-0.0021 \pm 0.0005$), ($0.0200 \pm 0.0003$, $-0.0024 \pm 0.0004$),
($0.0190 \pm 0.0005$, $-0.0019 \pm 0.0004$) respectively, showing that
there is no significant change in the interstellar and instrumental
polarization.}
\end{figure}

\begin{figure}
\includegraphics[angle=0,width=12cm]{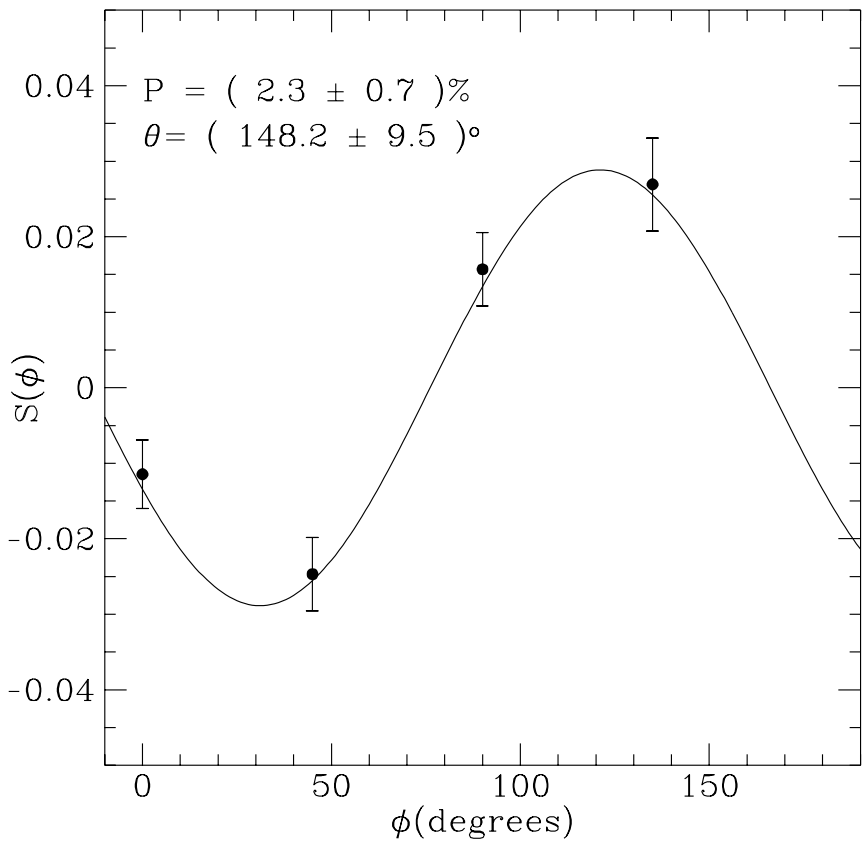}
\caption{\label{figure:cosfit}The parameter $S(\phi)$ at four different position angles $\phi$. The data are fit with a cosine function. The amplitude of the fit corresponds to the degree of linear polarization, and its maximum gives the position angle of the polarization.}
\end{figure}

\onecolumn

\begin{figure}
\includegraphics[angle=-90,width=12cm]{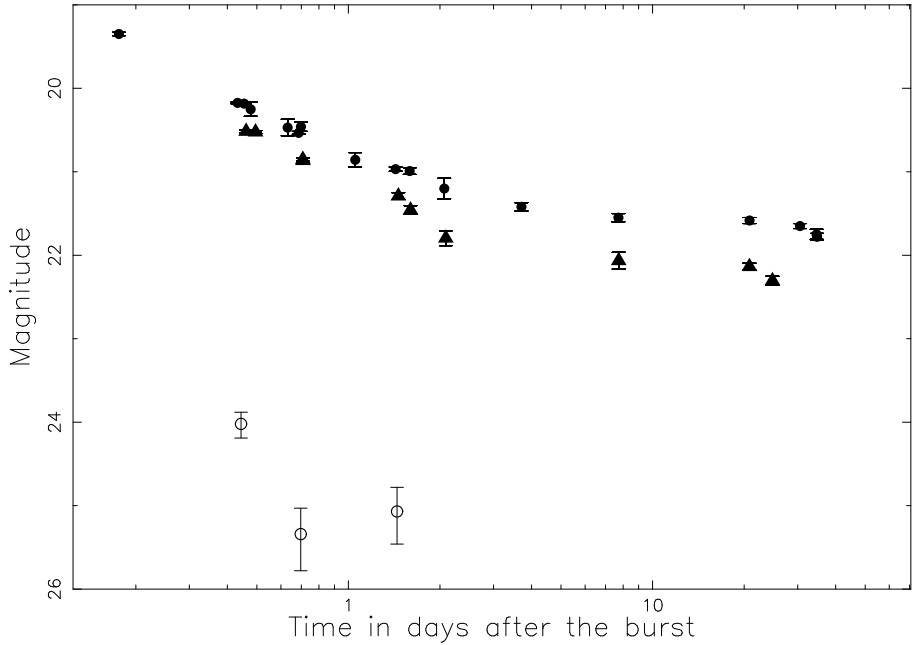}
\caption{\label{figure:lc-pol}The light curve of GRB\,990712. V and R
band data (filled circles and triangles respectively) are from
\citet{sahu00}. The open circles indicate the polarization magnitudes.}
\end{figure}

\begin{figure}
\includegraphics[angle=-90,width=15cm]{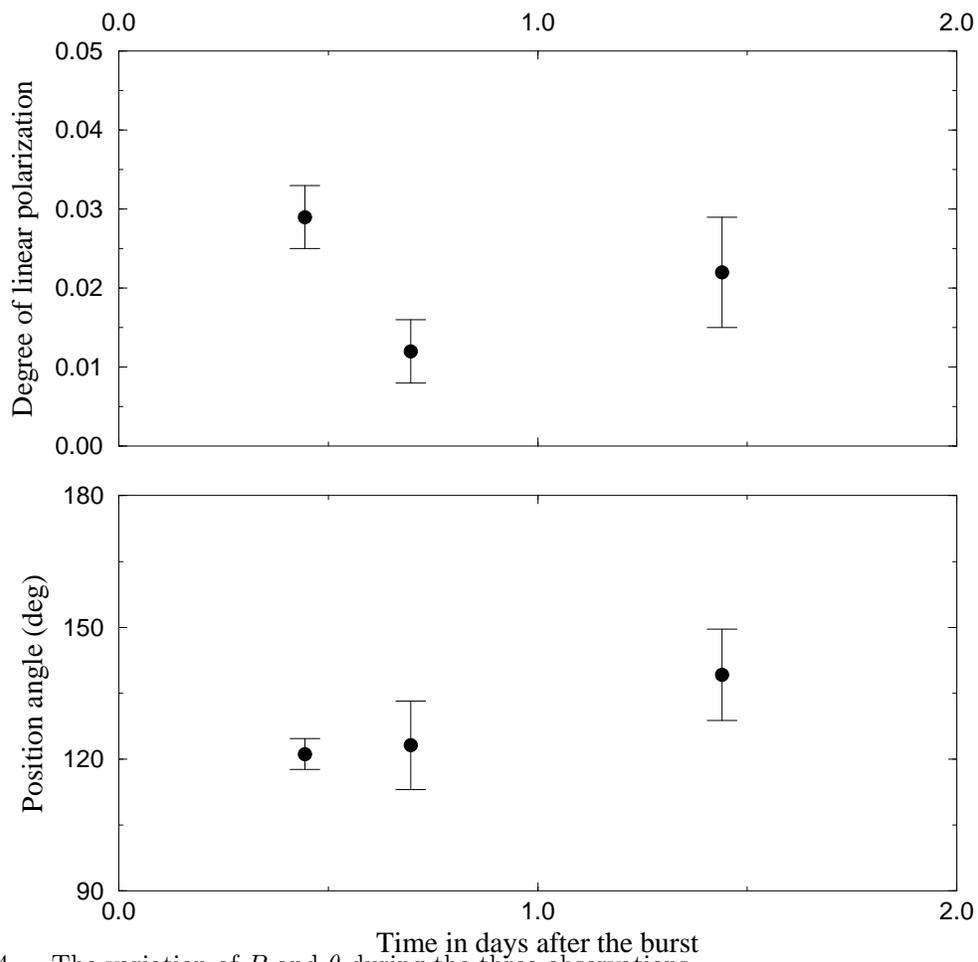}
\caption{\label{figure:ptheta-plot}The variation of $P$ and $\theta$
during the three observations.}
\end{figure}

\begin{deluxetable}{ccccccc}
\tabletypesize{\footnotesize}

\tablecaption{\label{table:log}Log of the polarimetric observations of
the GRB\,990712 afterglow with ESO VLT Antu (UT1) and FORS1. The
exposure time for epoch 1 and 2 was 4 $\times$ 300 seconds and for the
3rd epoch 4 $\times$ 450 seconds. $P_{av}$ and $\theta_{av}$ denote
the average values of the field stars (not corrected for instrumental
and interstellar polarization).} 

\tablehead{\colhead{UT date\tablenotemark{a}} & \colhead{Days} &
\colhead{seeing\tablenotemark{b}} & \colhead{$P$} & \colhead{$\theta$}
& \colhead{$P_{av}$} & \colhead{$\theta_{av}$} \\ \colhead{(1999)} &
\colhead{after the burst} & \colhead{(arcsec)} &
\colhead{(percentage)} & \colhead{(deg)} & \colhead{(percentage)} &
\colhead{(deg)}}

\startdata 
July 13.1402 & 0.4436 & 1.0 & $2.9 \pm
0.4$ & $121.1 \pm 3.5$ & $1.89 \pm 0.02$ & $175.5 \pm 0.3$\\ July
13.3936 & 0.6971 & 0.8 & $1.2 \pm 0.4$ & $116.2 \pm 10.1$ & $2.06 \pm
0.02$ & $176.3 \pm 0.2$ \\ July 14.1415 & 1.4450 & 1.1 & $2.2 \pm 0.7$
& $139.2 \pm 10.4$ & $1.91 \pm 0.02$ & $175.4 \pm 0.2$ \\ 
\enddata

\tablenotetext{a}{Mid-exposure date.}  

\tablenotetext{b}{The seeing varied somewhat during the measurements;
the values listed here are averages.}

\end{deluxetable}

\newpage

\end{document}